# Misuse of Antibiotics in Poultry Threatens Pakistan Community's Health

Muhammad Hamza[1], Hafeez Ur Rehman Ali Khera[2], Muhammad Umair Waqas[3] Ayesha Muazzam[4], Sania Tariq[5], Zain Kaleem[6], Waseem Akram[7], M Talha Mumtaz[8], Shehroz Ahmad[9], and Abdul Samad[10*]

**Affiliation Address:**

[1,2,3,4,5,6,7,8,9] Faculty of Veterinary and Animal Sciences MNS University of Agriculture, 25000, Multan, Pakistan

[10*] Department of Animal Sciences, Gyeongsang National University South Korea.

**Corresponding Author***

Abdul Samad samad@gnu.ac.kr

## Abstract

A survey was conducted from February 2022 to May 2022 on the usage of antibiotics at a poultry farm in different areas of Multan, Punjab Pakistan. A well-organized questionnaire was used for the collection of data. Sixty poultry farms were surveyed randomly in the Multan district. All of these Farms were using antibiotics. Antibiotics are commonly used for the treatment of diseases. Some are used as preventive medicine and a few are used as growth promotors. neomycin, erythromycin, oxytetracycline, streptomycin, and colistin are the broad-spectrum antibiotics that are being used commercially. Enrofloxacin and Furazolidone are the common antibiotics that are being used in Studies these days. The class of Fluoroquinolones is commonly used in poultry farms. Thirty-three patterns of antibiotic usage were observed at poultry farms. multi-drug practices were also observed on various farms. In this study, 25% of antibiotics are prescribed by the veterans while more than 90 % were acquired from the veterinary store. This study provides information about the antibiotics which are commonly being used in the study location district Multan. It is expected that the finding of this survey will be helpful in the development of new strategies against the misuse of antibiotics on farms.

**Introduction**

There is a worldwide movement to support a sustainable agriculture system. This is expected to be achieved by upgrading the farm practice, which will increase the profit and prove to be good for the environment. As it will provide antibiotics-free products. It is a great challenge to raise the broiler without the use of antibiotics. The production of broilers without using antibiotics is a great challenge in developing countries such as Pakistan, where antibiotics are being used frequently [1]. A broiler is commonly reared for meat production. It has soft meat, good tenderness, and low amounts of fat. The broiler has a very short period of production and has a good place globally as a meat bird. It provides food security and creates employment [2]. So sustainable broiler production depends upon both human and bird welfare. Which also involves the protection of the environment. Consumers are additionally concerned with the resistance of antibiotics in poultry. It is very crucial to produce broiler meat without using antibiotics. Because it will save poultry and other meat-producing animals from antimicrobial resistance which will promote public health.

There has been exponential growth in broiler meat production for global consumption and for generating profit. The low cost of production and the rapid progress of the economy are significant causes for the broiler industry expansion [3]. Although poultry farming is a very profitable business it also suffers from many challenges these challenges include infectious and non-infectious diseases which are caused by poor biosecurity and bad farm practices. Over the last few decades, some antibiotics are being used in poultry feed extensively. These are used to treat diseases and boost the immune system of poultry so that they can increase their performance and feed efficacy [4, 5]. The inappropriate use of antibiotics causes resistance in the birds. The most common pathway for the bacteria to get the resistance is by the genetic components such as a plasmid, necked DNA, and transposons. The gene transfer by the plasmids accelerates the process of bacteria gaining antimicrobial resistance [6]. The resistance in the bacteria can also be developed by the mutation in the chromosomes, as it occurs in the Fluoroquinolones resistance. Bacterial populations mutate by using the normal type of genetic variation which make antibiotics ineffective for them [7]. Bacterial resistance is developed by the interaction of the antibiotic and bacteria which leads to the instant removal of the sensitive type of bacteria and resistant types of bacteria are selected as a result. This distribution of resistance is a very complicated matter, and it creates a severe type of public health issue [8]. Influenza-like infections or any other type of viral infection facilitate the process of microbial resistance in poultry [9].

In Pakistan, poultry farmers extensively use antibiotics without the advice of veterinary advisors. It is being done on a commercial scale without following any guidelines [10]. The lack of knowledge and absence of proper veterinarians and the motive of the high profit led the farmers to use more times antibiotics than the normal limit [11]. To increase the overall growth performance of broilers, Antibiotics are used as a growth promotor with other growth promotor medicines [12, 13]. In consumers' residues of antimicrobial agents are stored in the various parts of the body such as tissues [14] These antibiotic residues mostly stay in the kidney, thigh liver, and breast meat of poultry. The availability of the farmers to these antibiotics without any prescription from specialized veterinarians results in antimicrobial resistivity in the farm animals [11]. Antibiotic resistance was present in nature already, but the rapid use of antibiotics has created the widespread pathogenic bacteria which are emerged recently [15]. As the same antibiotics which are being used in animals are used in humans so there is a great fear of antibiotic resistance in humans which is a severe concern for public health [16]. In the community, the resistant bacteria can spread very rapidly among humans and farm animals. So, there should be some new approaches used in broiler farming that can be marketed and could increase sustainably. Therefore, this research article highlights the recent scenario of the antibiotics which are being used in broiler production its global challenges, prospect, and approach for antibiotic-free productivity particularly in the context of Pakistan.

**Literature Review**

In Saudi Arabia 23 haphazardly chosen poultry homesteads and all veterinary drug stores in the Eastern Province of Saudi Arabia were overviewed for the anti-microbials utilized or apportioned. Further, a complete writing review was performed. 29 antimicrobial specialists were distinguished as being accessible for poultry use, of which 22 (75.9%) were significant for the treatment of human diseases. Enrofloxacin, oxytetracycline, ampicillin, neomycin, sulphamethoxazole, colistin, doxycycline, and erythromycin were the most often utilized drugs. Food-bone touchiness responses and the rise of microbial opposition, as well as cross-protection from the different gatherings of anti-infection agents in creatures and their exchange to human microorganisms, are irrefutable. The unreasonable utilization of anti-toxins in the poultry business can cause genuine medical problems this might influence general well-being since it will confuse the therapy of the contamination in people [17]. Thus, the veterinarians ought to boycott those anti-toxins which are being utilized in the two creatures and in people to forestall abiotic opposition it is enthusiastically suggested that there ought to be a division laid out by the public authority which will worry with the food and medications security which is strongly suggested nowadays.

The commonly used antibiotics in poultry can be reduced by using the proper farm practices, sanitization of the farm, and education of the farmers on the proper use of the antibiotics there should be veterinarians and proper enforcement of law and regulations which will reduce the use of the antibiotics on farms in Pakistan [18]. Broiler farmers are making use of antibiotics in farming against diseases which are also helpful in its better growth or production. But antibiotic residues in meat may cause antibiotic resistivity in consumers. There is a need to switch to alternatives to antibiotics. There are various difficulties we face in applying it. Considering consumer health by providing a safe product (Broiler) we need to overcome these difficulties.

In Sudan, anti-microbials can be gotten from drug stores without a remedy, bringing about an elevated degree of self-medicine [19] and likely anti-toxin opposition. It ought to be noticed that similar anti-infection agents utilized in human medications are utilized in domesticated animals. A few examinations have shown the spread of anti-microbial safe food-borne pathogenic microscopic organisms, for instance, Salmonella, Campylobacter, and Escherichia coli [20]. For instance, a review from Khartoum has uncovered various protections from ordinarily utilized anti-microbials against microorganisms causing mastitis among cows. One more review has announced separation and recognizable proof of safe Salmonella from chickens.

**Materials and methods**

A Cross-sectional survey was conducted from February to May 2022. This Survey was conducted using a well-organized questionnaire. A random survey was carried out at 60 farms in the district Multan. Information related to antibiotics such as the source of antibiotics, prescription, and the reason for using antibiotics on farms from the past year was documented. Information was gathered from the farm supervisor and interviews were also carried out on the spot. The poultry farms which were using more than two or three antibiotics at the same time were documented as multi-drug usage farms and the data for the past year was collected. These antibiotics are commonly prepared broad-spectrum antibiotics which contain two or different types of antibiotics.

**Data analysis**

The complete data was collected and then arranged in an excel sheet. All the responses which are collected from the questionnaire were presented in simple frequencies.

**Result**

A total of 60 farms were visited and the majority of them commercial-scale farms with a capacity of 30000 birds. 60(100%) of these farms were using more than one antibiotic at a time, 21(36.2%) were using antibiotics for therapeutic purposes, 17(29.3%) were used for prophylactic purposes, 19(32.8%) were using antibiotics for both purposes while 6 (6.9%) were using antibiotics as growth premotor. 29(50%) of farms indicated that those antibiotics were prescribed by a veterinary doctor, 25(43.1%) were self-medication practices and 6 (6.9%) were prescribed by animal health workers. 53(91,4%) antibiotics sources were from the antibiotics stores and 7 antibiotics were from the local sealers (Table 1). The usage pattern of antibiotics in a poultry farm in the past year is shown in Table 2. 33 patterns of antibiotics usage were observed in poultry farms, 41(70.7 %) farms were using more than one antibiotic at a time thus were multi-drug-using farm and 10(17.2%) farms were using one antibiotic during the past year. Quinolone was the common type of antibiotic used on farms. The commonly used antibiotics in the study area were mostly broad-spectrum antibiotics such as Neomycin, streptomycin, erythromycin, oxytetracycline, colistin, Furazolidone, and Enrofloxacin (Table 3).

**Table 1: Antibiotics usage patterns in some selected broiler farms in District Multan**

Total No of Farm visited – 60

| Description of items | Responses | Frequency of Antibiotics used | Percentage of Antibiotics used |
|---|---|---|---|
| Usage of Antibiotic on farms | Yes | 60 | 100% |
| | Veterinary doctors | 29 | 50.0% |

| sources of prescriptions | Self-medications | 25 | 43.1% |
| --- | --- | --- | --- |
| | Animal health worker | 6 | 6.9% |
| Reasons for antibiotic usage | Treatments | 21 | 36.2% |
| | Prophylaxis | 17 | 29.3% |
| | Treatment and the Prophylaxis | 19 | 32.8% |
| | Growth promotors | 06 | 6.9% |
| Sources of Antibiotic | Veterinary stores | 53 | 91.4% |
| | Local Vendors | 07 | 8.6% |

**Table 2: History of antibiotics used in poultry farms in the past year**

| Serial no | Commonly Used Antibiotics | N0 of farms |
| --- | --- | --- |
| 1. | Enrofloxacin's, Gentamycin's, Furazolidones | 02 |
| 2. | Oxytetracyclines, NeocerylR | 02 |
| 3. | Colistin's, Flumequines | 01 |
| 4. | Doxycycline's, Gentamycin's | 01 |
| 5. | Tylosin's, Furazolidones, NeocerylR | 01 |
| 6. | NCOs, Streptomycin's, Doxycycline's | 01 |
| 7. | Ciprofloxacin's | 01 |
| 8. | Ciprofloxacin's, Penicillin's, NeocerylR | 02 |
| 9. | Tylosin's, Oxytetracyclines, Enrofloxacin's | 01 |
| 10. | Gentamycin's, Ciprofloxacin's | 01 |
| 11. | Enrofloxacin's | 04 |
| 12. | Enrofloxacin's, Furazolidones | 01 |
| 13. | Tylosin's, Norfloxacin's, Flumequines | 01 |
| 14. | Gentamycin's, Doxycycline's, Enrofloxacin's, Norfloxacin's | 01 |
| 15. | NCOs, NeocerylR | 02 |
| 16. | Ciprofloxacin's, NCOs, Enrofloxacin's, Furazolidones, Penicillin's | 01 |
| 17. | Furazolidones | 01 |

| 18. | Gentamycin's, Streptomycin's, Enrofloxacin's | 01 |
|---|---|---|
| 19. | NeocerylR, Penicillin's | 02 |
| 20. | Furazolidones, Colistin's, Enrofloxacin's | 02 |
| 21. | Ciprofloxacin's, Furazolidones, Gentamycin's, Tylosin's, Neoceryl R | 01 |
| 22. | Gentamycin's, Enrofloxacin's, Neoceryl R | 01 |
| 23. | NCO, Enrofloxacin's | 01 |
| 24. | Gentamycin's, Doxycycline's, Neoceryl R | 01 |
| 25. | NeocerylR | 10 |
| 26. | Neomycin, Oxytetracycline | 02 |
| 27. | Oxytetracyclines | 02 |
| 28. | NCOs | 01 |
| 29. | Tylosin's, Enrofloxacin's | 01 |
| 30. | Tylosin's, Neoceryl R | 04 |
| 31. | Enrofloxacin's, Oxytetracyclines, Gentamycin's, Penicillin s | 01 |
| 32. | Penicillin's, Streptomycin's, Norfloxacin's | 02 |
| 33. | Penicillin's, Streptomycin, Tetracyclines | 02 |
| 34. | Total No | 60 |

**Table 3 Commonly used antibiotics in selected poultry farms of District Multan**

| Antibiotics Used | Frequency of Antibiotics | Percentages |
|---|---|---|
| 1. Enrofloxacin's | 16 | 27.60 |
| 2. Gentamycin's | 10 | 17.20 |
| 3. NeocerylR | 21 | 36.20 |
| 4. Furazolidones | 12 | 20.70 |
| 5. Colistin's | 03 | 5.20 |
| 6. Penicillin's | 09 | 15.50 |
| 7. Ciprofloxacin's | 05 | 8.60 |
| 8. Norfloxacin's | 03 | 5.20 |
| 9. Tylosin's | 09 | 15.50 |
| 10. NCO's | 06 | 10.30 |

| 11. Oxytetracyclines | 06 | 10.30 |
| 12. Doxycyclinene's | 05 | 8.60 |
| 13. Streptomycin's | 05 | 8.60 |
| 14. Tetracyclines | 02 | 1.70 |
| 15. Flumequines | 03 | 3.50 |

**Discussion**

From the survey, it was disclosed that almost all poultry farms used antibiotics for the purpose of treatment of disease, its prevention, and also for increased growth in birds. Previous reports revealed the same findings. [19,20 ,21,22]. Few poultry farms relied on antibiotics for growth enhancement. It is at odds with Olatoye who revealed that 86% of poultry farms in Ibadan, Nigeria, used antibiotics for growth enhancement. [23] findings also revealed the increased usage of antibiotics in poultry farms for preventive treatment and growth in birds. An increase in antibiotics usage may be due to poor sanitation, biosecurity, and lack of proper nutrition which resulted in its exposure to various bacteria and other disease-causing agents which contributed to spreading diseases.

The effects of antimicrobial use in livestock production are reported [24, 25]. We observed in this study, that the high use of antibiotics in animal source foods can affect the human food chain in form of drug residues. It can lead to a lower effect on those drugs which can be used for treating human disease. Additionally, the use of antimicrobials may be selected for antibiotic-resistant bacterial strains. They can transmit resistance genes to other pathogenic and non-pathogenic bacteria. In this way, human beings are safe from bacterial resistance after using poultry products.

The use of antibiotics prevents infections in poultry sheds but overuse of antibiotics is harmful to human health. As they create resistance in the body. The veterinary department should make the rule and qualified staff are recommended in the sheds. The misuse of antibiotics in local sheds should be prevented. An alternative such as the use of vaccines and sanitation of the environment would decrease the use of antibiotics.

## Conclusions and Recommendations

The main reason for misuse of antibiotics in District Multan is their misapprehension, their poor awareness, ignorance of it, and easy reach to antibiotics. There is a dire need for proper education and carrying out strict actions to prevent the use of antibiotics. Strict policies on antibiotics prescription and their use should be made. Farmers must be properly made aware of the issue of antibiotic resistance. As the common farming in Khartoum is subsistence, studies should include subsistence farmers.

## Acknowledgment

We appreciate all the team members, poultry farm managers, supervisors interviewed, and especially Muhammad Hamza and Abdul Samad for their time and efforts.

## References

1. Masud, A.A.; Rousham, E.K.; Islam, M.A.; Alam, M.U.; Rahman, M.; Mamun, A.A.; Sarker, S.; Asaduzzaman, M.; Unicomb, L.Drivers of Antibiotic Use in Poultry Production in Bangladesh: Dependencies and Dynamics of a Patron-Client Relationship. Front. Vet. Sci. 2020, 7, 78. [CrossRef] [PubMed]

2. Kryger, K.N.; Thomsen, K.; Whyte, M.; Dissing, M. Smallholder Poultry Production: Livelihoods, Food Security and Sociocultural Significance; Series FAO Smallholder Poultry Production; FAO: Rome, Italy, 2010; p. 4.

3. Najeeb, A.; Mandal, P.; Pal, U. Efficacy of fruits (red grapes, gooseberry, and tomato) powder as natural preservatives in restructured chicken slices. Int. Food Res. J. 2014, 21, 2431–2436.

4. Tollefson, L.; Miller, M.A. Antibiotic use in food animals: Controlling the human health impact. J. AOAC Int. 2000, 83, 245–254. [CrossRef] [PubMed]

5. Gaskins, H.; Collier, C.; Anderson, D. Antibiotics as growth promotants: Mode of action. Anim. Biotechnol. 2002, 13, 29–42. [CrossRef] [PubMed]

6. Davies, J.; Davies, D. Origins and evolution of antibiotic resistance. Microbiol. Mol. Biol. Rev. 2010, 74, 417–433. [CrossRef]

7. Landers, T.F.; Cohen, B.; Wittum, T.E.; Larson, E.L.A Review of Antibiotic Use in Food Animals: Perspective, Policy, and Potential. Public Health Rep. 2012, 127, 4–22. [CrossRef]

8. Diaz-Sanchez, S.; Moscoso, S.; Solís de los Santos, F.; Andino, A.; Hanning, I. Antibiotic use in poultry; A driving force for organic poultry production. Food Prot. Trends 2015, 35, 440–447.

9. Marshall, B.M.; Levy, S.B. Food Animals and Antimicrobials: Impacts on Human Health. Clin. Microbiol.Rev. 2011, 24, 718–733. [CrossRef]

10. Hoque,R.; Ahmed, S.M.; Naher,N.; Islam, M.A.; Rousham, E.K.; Islam, B.Z.; Hassan,S. Tackling Antimicrobial Resistance in Bangladesh: A Scoping Review of Policy and Practice in Human, Animal and Environment Sectors. PLoS ONE 2020, 15, e0227947. [CrossRef]

11. Saiful, I.K.B.M.; Shiraj-Um-Mahmuda, S.; Hazzaz-Bin-Kabir, M. Antibiotic Usage Patterns in Selected Broiler Farms of Bangladesh and their Public Health Implications. J. Public Health Dev. Ctries. 2016, 2, 276–284.

12. Stutz, M.W.; Lawton,G.C.Effects of diet and antimicrobials on growth, feed efficiency, intestinal Clostridium perfringens, and ileal weight of broiler chicks. Poult. Sci. 1984, 63, 2036–2042. [CrossRef] [PubMed]

13. Sirdar, M.M.; Picard, J.; Bisschop, S.; Gummow, B. A questionnaire survey of poultry layer farmers in Khartoum State, Sudan, to study their antimicrobial awareness and usage patterns. Onderstepoort J. Vet. Res. 2012, 79, 1–8. [CrossRef] [PubMed]

14. Sattar, S.; Hassan, M.M.; Islam, S.K.M.A.; Alam, M.; Faruk, M.S.A.; Chowdhury, S.; Saifuddin, A.K.M. Antibiotic residues in broiler and layer meat in Chittagong district of Bangladesh. Vet. World 2014, 7, 738–743. [CrossRef]

15. D'Costa, V.M.; King, C.E.; Kalan, L.; Morar, M.; Sung, W.W.L.; Schwarz, C.; Froese, D.; Zatula, G.; Camels, F.; Debruyme, R.; et al. Antibiotic resistance is ancient. Nat. Lett. 2011, 477, 457–461. [CrossRef] [PubMed]

16. Laxminarayan, R.; Van Boeckel, T.; Teillant, A. The Economic Costs of Withdrawing Antimicrobial Growth Promoters from the Livestock Sector. OECD Food Agric. Fish. 2015. [CrossRef]

17. Zaki H. Al-Mustafa, PhD; Mastour S. Al-Ghamdi, PhD. Use of Antibiotics In The Poultry Industry In Saudi Arabia: Implications For Public Health Ann Saudi Med 2002;22(l-2):4-7.

18. LindonneGlasgow ,1 MartinForde ,1 DarrenBrow,1 CatherineMahoney,1 StephanieFletcher ,2 andShellyRodrigo1. Antibiotic Use in Poultry Production in Grenada Veterinary Medicine International Volume 2019, Article ID 6785195, 7 pages

19. Aarestrup FM (1999). Association between the consumption of antimicrobial agents in animal husbandry and the occurrence of resistant bacteria among food animals. International Journal of Antimicrobial Agents, 12(4): 279-85.

20. Van Duijkeren E & Houwers DJ (2000). A critical assessment of antimicrobial treatment in uncomplicated Salmonella enteritis. Veterinary Microbiology, 73(1): 61-75

21. Zhao S, McDermott PF & Friedman S (2006). Antimicrobial resistance and genetic relatedness among Salmonella from retail foods of animal origin: NARMS retail meat surveillance. Food-borne Pathogen and Diseases, 3(1): 106-117

22. Alo OS & Ojo O (2007). Use of antibiotics in food animals; A case of a major veterinary outlet in Ekiti State, Nigeria. Nigerian Veterinary Journal, 28(1): 80-82. Olatoye IO (2011). Antibiotics use and resistance patterns of Salmonella species in poultry from Ibadan, Nigeria. Tropical Veterinarian, 29(1): 28-35.

23. Sirdar MM, Picard J, Bisschop S & Gummow B, (2012). A questionnaire survey of poultry layer farmers in Khartoum State, Sudan, to study their antimicrobial awareness and usage patterns', Onderstepoort Journal of Veterinary Research 79(1): 8 pages

24. Van den Bogaard AE & Stobberingh EE (2000). Epidemiology of resistance to antibiotics. Links between animals and humans. International Journal of Antimicrobial Agents. 14(4): 327-33.

25. Molla B, Alemayehu D & Salah W (2003). Sources and distribution of Salmonella serotypes isolated from food animals, slaughterhouse personnel and retail meat products in Ethiopia 1997-2002. Ethiopian Journal of Health Development, 17(1): 63-70.